\documentclass{aa}
\usepackage{graphics}

\begin{document}      

   \title{NGC~4254: a spiral galaxy entering the Virgo cluster}

   \author{B.~Vollmer\inst{1,2}, W.~Huchtmeier\inst{2} \& W.~van Driel\inst{3}}

   \offprints{B.~Vollmer, e-mail: bvollmer@astro.u-strasbg.fr}

   \institute{CDS, Observatoire astronomique de Strasbourg, UMR 7550, 11, rue de
              l'universit\'e, 67000 Strasbourg, France \and
              Max-Planck-Institut f\"ur Radioastronomie, Auf dem H\"ugel 69,
              D-53121 Bonn, Germany \and
	      Observatoire de Paris, Section de Meudon, GEPI, CNRS UMR 8111 and 
	      Universit\'e de Paris 7, 5 place Jules Janssen, 92195 Meudon, Cedex, France.
              }

   \date{Received / Accepted}

   \titlerunning{NGC~4254}

\abstract{
Deep Effelsberg H{\sc i} spectra of the one-armed, bright Virgo cluster spiral galaxy NGC~4254 are presented.
Five different positions were observed in the 21~cm H{\sc i} line with the Effelsberg 100-m telescope: 
one in the center and 4 located one HPBW to the NE, NW, SW, and SE, respectively, from the galaxy center.
The spectra are compared to existing deep VLA observations, and the single dish and
interferometric H{\sc i} data are used to constrain a dynamical model which includes the effects
of ram pressure. The peculiar, one-armed spiral pattern of NGC~4254 and its distorted and kinematically 
perturbed atomic gas distribution
can be explained by a close and rapid encounter $\sim 280$~Myr ago with another massive Virgo galaxy,
followed by ram pressure stripping that is still ongoing.
The stripping occurs almost face-on, since the angle between the disk and the orbital plane is $70^{\circ}$.
The galaxy with which NGC~4254 had its encounter is tentatively identified as the lenticular NGC~4262.
}

\maketitle

\section{Introduction}

NGC~4254 is a bright spiral galaxy which is located at the periphery of the Virgo cluster,
at a projected distance of $3.7^{\circ}$ ($\sim 1$~Mpc\footnote{We use $D=17$~Mpc for the distance 
to the Virgo cluster.}) from the cluster center (M87). Its basic parameters are listed in 
Tab.~\ref{tab:parameters}. The spiral structure of NGC~4254 shows an important
m=1 mode, giving it a one-armed appearance (Fig.~\ref{fig:stars}). 
The m=2 mode, which is most prominent in most grand-design spirals, is almost
absent. Iye et al. (1982) suggested that the intrinsic properties of NGC~4254 are responsible
for the predominant m=1 mode.

Phookun et al. (1993) obtained deep H{\sc i}  21~cm line observations of this galaxy
with the VLA.
These deep observations revealed, in addition to the galactic H{\sc i} disk component,
H{\sc i} clouds superposed on and beyond the H{\sc i} disk with velocities that
do not follow the disk rotation pattern. They also found very low surface density
atomic gas up to $\sim 6'$ ($\sim 30$~kpc) to the north-west of the galaxy center.
Based on their observations, Phookun et al. (1993) suggested that the low surface density
gas represents a remnant of an entity that was tidally disrupted by NGC~4254 and is now merging
with it. In this scenario the m=1 spiral structure was stimulated by the infalling gas.

\begin{table}
      \caption{Physical Parameters of NGC~4254}
         \label{tab:parameters}
      \[
         \begin{array}{lr}
            \hline
            \noalign{\smallskip}
        {\rm Other\ names} &  {\rm M~99} \\
                & {\rm VCC~307} \\
                & {\rm CGCG~098-144}  \\
        $$\alpha$$\ (2000)$$^{\rm a}$$ &  12$$^{\rm h}18^{\rm m}49.6^{\rm s}$$\\
        $$\delta$$\ (2000)$$^{\rm a}$$ &  14$$^{\circ}24'59''$$\\
        {\rm Morphological\ type}$$^{\rm a}$$ & {\rm Sc} \\
        {\rm Distance\ to\ the\ cluster\ center}\ ($$^{\circ}$$) & 3.7\\
        {\rm Optical\ diameter\ D}_{25}$$^{\rm a}$$\ ($$'$$) & 5.4\\
        {\rm B}$$_{T}^{0}$$$$^{\rm a}$$ & 10.44\\ 
        {\rm Systemic\ heliocentric\ velocity}$$^{\rm a}$$\ {\rm (km\,s}$$^{-1}$$)\ & 2407$$\pm$$3\\
        {\rm Distance\ D\ (Mpc)} & 17 \\
        {\rm Vrot}$$_{\rm max}\ {\rm (km\,s}$$^{-1}$$) & 150$$^{\rm b}$$ \\
        {\rm Major axis p.a.} & 68$$^{\circ}$$\ $$^{\rm b}$$\\
        {\rm Inclination\ angle} & 42$$^{\circ}$$\ $$^{\rm b}$$ \\
        {\rm HI\ deficiency}^{\rm c}$$ &  0.17$$\pm$$0.2\\
        \noalign{\smallskip}
        \hline
        \end{array}
      \]
\begin{list}{}{}
\item[$^{\rm{a}}$] RC3, de Vaucouleurs et al. (1991)
\item[$^{\rm{b}}$] Phookun et al. (1993)
\item[$^{\rm{c}}$] Cayatte et al. (1994)
\end{list}
\end{table}

NGC~4254 resembles in many respects NGC~4654 (Phookun \& Mundy 1995), 
another bright Virgo cluster spiral galaxy that
is located at about the same projected distance from the cluster center, but at the opposite side of the 
cluster, since both galaxies (i) show a prominent m=1 spiral structure, (ii) are not 
H{\sc i} deficient, and (iii) have an extended, very low surface density gas tail.

Vollmer (2003) showed that the m=1 spiral structure of NGC~4654 can be explained by
a tidal interaction of NGC~4654 with its companion NGC~4639 that took place $\sim 500$~Myr ago,
while its extended gas tail can be explained by ongoing ram pressure 
stripping (Gunn \& Gott 1972). Since the outer part of the gas disk was perturbed by the gravitational
interaction, only a small amount of ram pressure is needed to create the gas tail.
Thus, only the combination of a close encounter with NGC~4639 followed by continuing
moderate ram pressure can reproduce all observed properties of NGC~4654.
In this article we investigate if the same evolutionary scenario, involving a close
encounter and ongoing ram pressure stripping applies to NGC~4254.

The Effelsberg observations are described in Sect.~\ref{sec:obs}, followed by the results
in Sect.~\ref{sec:results}. The model is introduced in Sect.~\ref{sec:model}. We present two
different simulations in Sect.~\ref{sec:simulation}: (i) a close and rapid encounter between 
NGC~4254 and another massive Virgo cluster galaxy (Sect.~\ref{sec:tidal}) and (ii) a close and rapid 
encounter together with ram pressure stripping (Sect.~\ref{sec:ram}). The simulations 
are compared to  H{\sc i} observations in Sect.~\ref{sec:compare} and discussed
in Sect.~\ref{sec:discussion}, followed by our conclusions in Sect.~\ref{sec:conclusions}.

\section{Observations \label{sec:obs}}

Between January and March 2002 we performed 21-cm line observations with the 100m Effelsberg
telescope at 5 different positions in and around NGC~4254 at a frequency 
corresponding to the systemic velocity of NGC~4254 and a bandwidth of 12.5~MHz. 
The two-channel receiver had a system noise of $\sim$30~K. The 1024 channel autocorrelator
was split into four banks of 256 channels each, yielding a
channel separation of $\sim$10~km\,s$^{-1}$. We further binned the channels to obtain
a final channel separation of $\sim$20~km\,s$^{-1}$. The central position of the galaxy and four 
positions at a distance of one HPBW (9.3$'$) to the NW, SW, SE, and
NE, respectively, from the galaxy center were observed in on--off mode
using pairs of 5~min on source and 5~min off source integrations. 
The total integration times are shown in Table~\ref{tab:table}.
\begin{table}
      \caption{Integration times and rms.}
         \label{tab:table}
      \[
         \begin{array}{lccccc}
           \hline
           \noalign{\smallskip}
           {\rm position} & {\rm Center} & {\rm Northwest} & {\rm Southwest} & {\rm Southeast} & {\rm Northeast} \\
	   \noalign{\smallskip}
	   \hline
	   \noalign{\smallskip}
	   $$\Delta t$$\ {\rm (min)} & 60 & 60 & 60 & 40 & 45 \\ 
           \noalign{\smallskip}
	   \hline
	   \noalign{\smallskip}
	   {\rm rms\ (mJy)} & 2.8 & 1.8 & 1.4 & 3.5 & 1.7 \\
	   \noalign{\smallskip}
        \hline
        \end{array}
      \]
\end{table}
Care was taken to avoid other Virgo galaxies with radial velocities within our bandwidth
in all 5 on and off source positions. We used 3C286 for pointing and flux
calibration. The on and off observation times ranged between 80~min and 120~min per position.
The resulting noise (Table~\ref{tab:table}), which is partly determined by small amplitude 
interferences, is close to the theoretical noise of 2~mJy per hour of integration.

\section{Results \label{sec:results}}

\begin{figure*}
        \resizebox{\hsize}{!}{\includegraphics{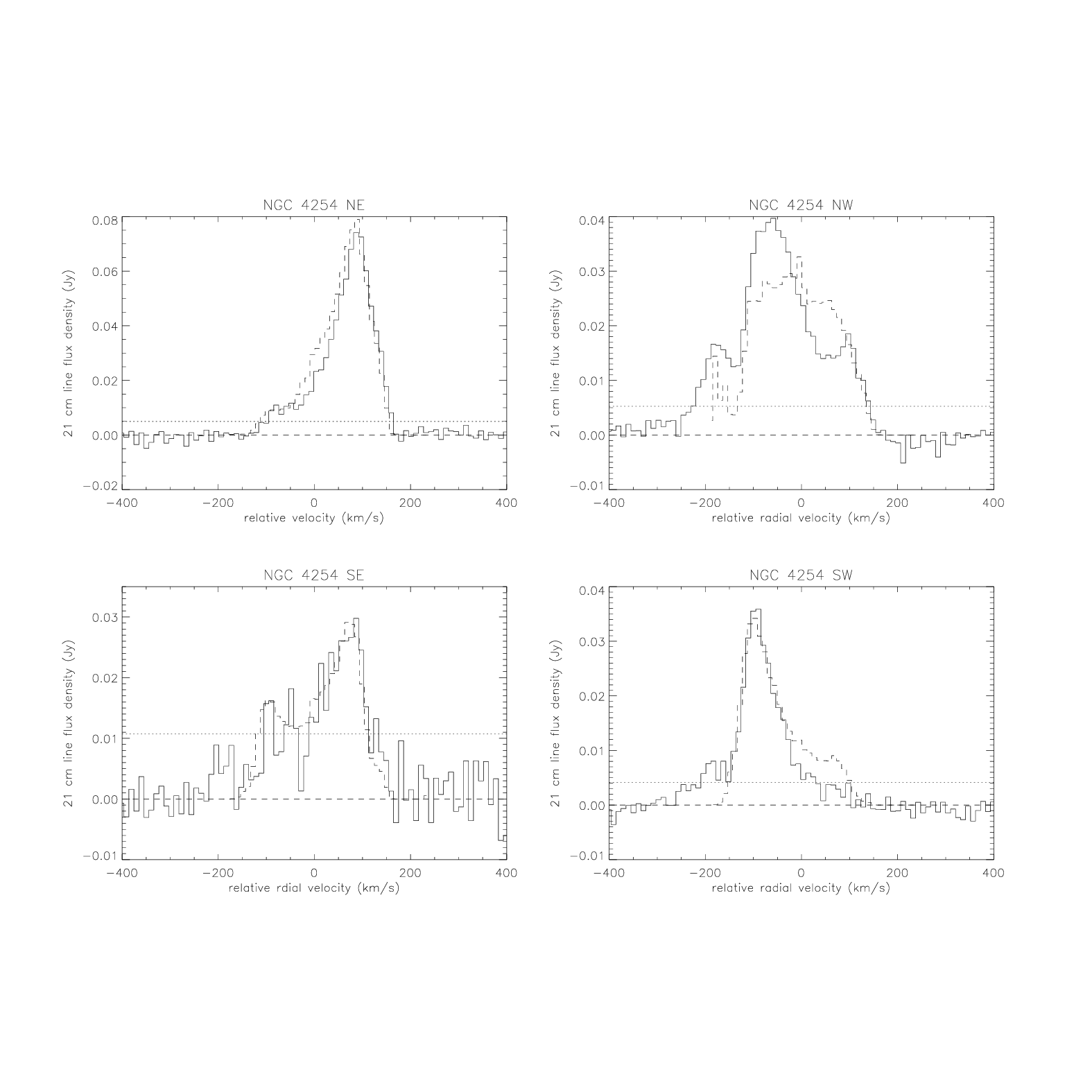}}
        \caption{Solid lines: Effelsberg 100-m spectra of the four off-center positions.
	Their locations with respect to the galaxy center are marked on top of each panel.
	Dashed line: synthesized VLA spectra which only show H{\sc i} disk emission. 
	Dotted line: 3$\sigma$ noise levels of the
	Effelsberg spectra. Velocities are given relative to the systemic velocity of NGC~4254,
	2407~km\,s$^{-1}$.
        } \label{fig:spectra}
\end{figure*}

The 5 Effelsberg spectra are shown in Fig.~\ref{fig:spectra} and Fig.~\ref{fig:cspectra}.
In order to compare our single-dish H{\sc i} spectra to interferometric data, where the
galaxy is spatially resolved, we use the VLA 21~cm C and D array data of Phookun et al. (1993). 
These data have a spatial resolution of $32'' \times 28''$ and a channel separation
of 10~km\,s$^{-1}$. The bulk of the H{\sc i} emission detected with the VLA is contained within
a $\sim 6' \times 5'$ area, while a faint blob of H{\sc i} was found $\sim 3'$ south-east
of the galaxy center and a very faint H{\sc i} tail was detected towards the north-west,
extending out to $10'$ from the center.

\begin{figure}
        \resizebox{8cm}{!}{\includegraphics{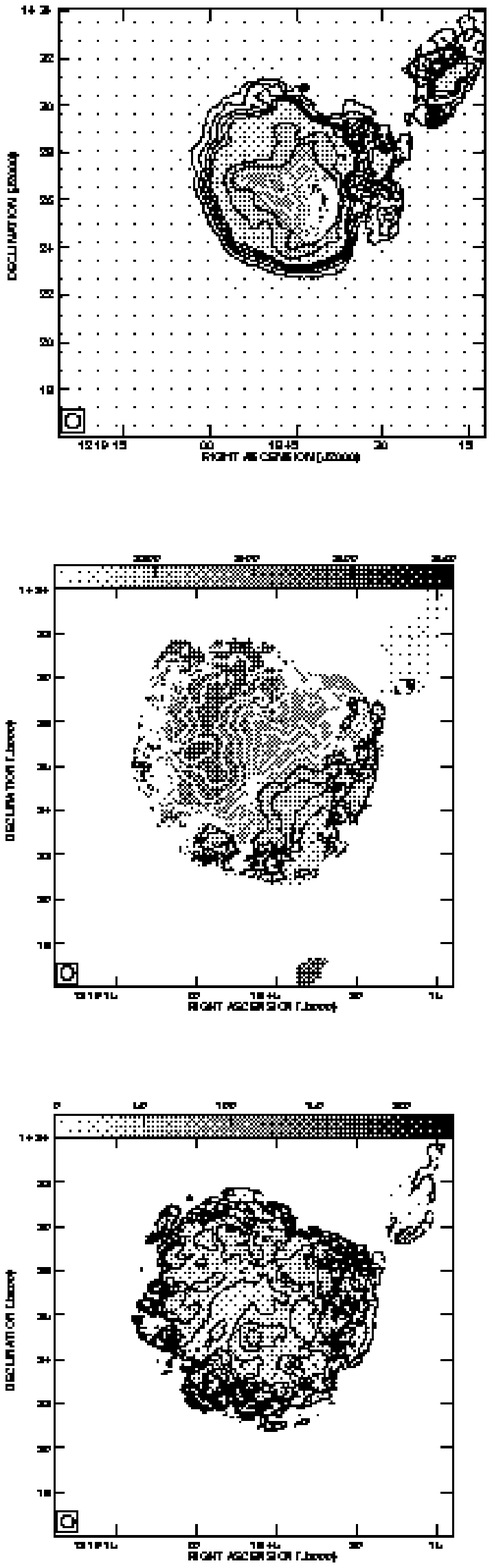}}
        \caption{H{\sc i} observations of NGC~4254 (Phookun et al. 1993).
	  Masks were produced from the data cube which was smoothed to a resolution 
	  of $4'$. The spectra were integrated within these masks.
	  Top panel: H{\sc i} distribution. The contour levels in all images are 
	  (0.02, 0.04, 0.06, 0.08, 0.1, 0.2, 0.3, 0.4, 0.5,
	  0.6, 0.7, 0.8, 0.9, 1.0) times the maximum of the image.
	  Middle panel: velocity field. The contour levels of all images are at 2250-2510~km\,s$^{-1}$, 
	  at intervals of $20$~km\,s$^{-1}$. Bottom panel: velocity dispersion.
	  The contour levels in all images are at 10-60~km\,s$^{-1}$, in steps
	  of $10$~km\,s$^{-1}$.
        } \label{fig:deep}
\end{figure}

In order to enhance the diffuse and extended H{\sc i} emission, we smoothed
the data cube to a resolution of $2'$, $3'$, $4'$, and $5'$. Three dimensional masks were produced
by setting all pixels to one where the intensity is above $5 \sigma$ and zero otherwise.
This mask was applied to the initial data cube.
For the 5 positions synthesized spectra from the VLA data were produced using the position of
the pointing and the beamsize of the Effelsberg 100-m telescope at 21~cm (9.3$'$).
We thus use the spatial information of the VLA data to construct the synthesized spectra.
It turned out that a smoothing to a $4'$ resolution leads to a maximum flux density
in the spectra. Therefore, all synthesized spectra are made with a $4'$ smoothing.
Moment maps were made of the masked cube without clipping (Fig.~\ref{fig:deep}). 
In these maps the north-western, very low surface density
plume is well detected.

The resulting
spectrum for the central position is represented by the dashed line in the lower panel of
Fig.~\ref{fig:cspectra}. We find a total single-dish flux of 67.7~Jy\,km\,s$^{-1}$, which corresponds
to an atomic gas mass of $4.6\,10^{9}$~M$_{\odot}$. The central velocity is 2406$\pm 10$~km\,s$^{-1}$,
the linewidths at 50\% and 20\% of the peak flux are $W_{50}=230$~km\,s$^{-1}$ and $W_{20}=270$~km\,s$^{-1}$.
The total and peak flux densities of Phookun et al. (1993) are within $\sim$10\% of our values, 
which is within the calibration accuracy. Both spectra agree reasonably well.

Fig.~\ref{fig:spectra} shows the 100-m spectra of the 4 off-center positions (solid lines)
together with the synthesized VLA spectra. 
The NE and SE Effelsberg and VLA spectra are in very good agreement.
The Effelsberg NW spectrum shows excess emission between $-100$ and $0$~km\,s$^{-1}$
and less emission between $0$ and $100$~km\,s$^{-1}$ with respect to the VLA
spectrum. The Effelsberg SW spectrum shows excess emission around $-200$~km\,s$^{-1}$
and less emission between $0$ and $100$~km\,s$^{-1}$ with respect to the VLA
spectrum. These differences can be explained by three effects: (i) pointing errors
of the Effelsberg telescope, (ii) disk emission detected by the sidelobes of
the Effelsberg telescope and (iii) extended ($ > 15'$) and/or very low surface density 
($ < 2 \times 10^{18}$~cm$^{-2}$) gas which has not been detected by the VLA. 
The excess emission of the NW and SW VLA spectra in the velocity range between
0 and 100~km\,s$^{-1}$ are possibly due to pointing errors of the Effelsberg telescope. 
A pointing error of about 10\,\% of the telescope's HPBW to the west can account for the 
VLA excess emission with respect to the Effelsberg spectrum. 
However, such a pointing error is unlikely, though in principle not impossible,
as it is 3 to 5 times the typical
pointing correction we had to apply during the observing run.
It is not excluded that the Effelsberg excess emission with respect to the VLA spectrum is due
to a sidelobe detection of the disk emission of the galaxy. 
The Effelsberg antenna diagram shows 4 first sidelobe maxima at a distance of $\sim 13'$ from the
beam center and at a -20~db level (1\,\% of the peak). One percent of the emission of the
central profile (Fig.~\ref{fig:cspectra}) is $\sim 3$~mJy, which corresponds approximately
to the $3\sigma$ noise levels indicated in the NW and SW spectra. 
Indeed, an inspection of the parallactic angle of the Effelsberg observations showed
that its variation was small for the eastern positions, where the galaxy was located near
a minimum of the sidelobes, whereas the variation was larger for the SW position and
maximum for the NW position where the maximum of the sidelobe passed over the galactic disk. 
As a conclusion, the differences between the western Effelsberg and VLA spectra can
in principle be explained by a mixture of effects (i), (ii) and (iii). However, we were not able
to sort out which effect dominates.
The north-eastern position only shows disk emission, whereas the south-western and south-eastern 
positions contain additional low surface density, low velocity gas (at $\sim -200$~km\,s$^{-1}$).
The most notable difference between the spectra occurs at the north-west position, where the
Effelsberg and VLA spectra show
three peaks: one due to galactic rotation at high velocities, a major peak at 
$-50$~km\,s$^{-1}$ and a third, distinct peak at $\sim -180$~km\,s$^{-1}$.
At the latter two velocities gas clumps were detected by Phookun et al. (1993) in the north-western tail.
Our spectrum reveals a bimodal velocity structure of the extended, low surface density
gas tail, which is already present, though less broad, in the VLA synthesized spectrum. 
We find a total line flux of 3.1~Jy\,km\,s$^{-1}$ ($2.1\,10^{8}$~M$_{\odot}$) or
4.6\% of the total H{\sc i} flux in the extended tail, more than the 3\% of the total flux found by
Phookun et al. (1993).

\section{The model \label{sec:model}}

Since the model we used is described in detail in Vollmer et al. (2001) and Vollmer (2003), 
we will only briefly summarize its main features here.
The particles represent gas cloud complexes which are 
evolving in an analytically given gravitational potential of the galaxy.

The non--collisional component of the NGC~4254 model consists of 49\,125 particles, which simulate
the galactic halo, bulge, and disk.
The characteristics of the different galactic components shown in
Tab.~\ref{tab:param} are the number of particles $N$, the particle mass $M$, and the
smoothing length $l$.
\begin{table}
      \caption{Parameters of the different galactic components.}
         \label{tab:param}
      \[
         \begin{array}{llll}
           \hline
           \noalign{\smallskip}
           {\rm component} & N & M\ ({\rm M}$$_{\odot}$$) & l\ ({\rm pc}) \\
           \hline
           {\rm halo} & 16384 & $$9.2\,10^{6}$$ & 1200 \\
           {\rm bulge} & 16384 & $$3.2\,10^{5}$$ & 180 \\
           {\rm disk} & 16384 & $$1.6\,10^{6}$$ & 240 \\
	   {\rm companion} & 11000 & $$9.2\,10^{6}$$ & 1200 \\
           \noalign{\smallskip}
        \hline
        \end{array}
      \]
\end{table}
The particle trajectories are integrated using an adaptive timestep for
each particle as described in Springel et al. (2001) and applying
the following criterion for an individual timestep:
\begin{equation}
\Delta t_{\rm i} = \frac{20~{\rm km\,s}^{-1}}{a_{\rm i}}\ ,
\end{equation}
where $a_{i}$ is the acceleration of particle i.
The minimum of all $\Delta t_{\rm i}$ defines the global timestep used 
for the Burlisch--Stoer integrator that integrates the collisional
component.

The initial conditions setup was made using the program described in
Hernquist (1993) with the following parameters:
\begin{itemize}
\item
halo: mass $M=1.5\,10^{11}$~M$_{\odot}$, core radius $r_{\rm c}$=3~kpc, tidal radius
$r_{\rm t}$=30~kpc, cutoff radius $R$=90~kpc,
\item
bulge: mass $M=5.2\,10^{9}$~M$_{\odot}$, scale length $l$=0.3~kpc, cutoff radius
$R$=30~kpc,
\item
disk: mass $M=2.6\,10^{10}$~M$_{\odot}$, scale length $l$=3~kpc, cutoff radius
$R$=45~kpc, disk thickness $z_{0}$=600~pc.
\end{itemize}

The collisional component of the model consists of 20\,000 model particles 
representing gas cloud complexes of different masses that are rotating within the gravitational 
potential of the galaxy. The total gas mass is $M_{\rm gas}^{\rm tot}=5.8\,10^{9}$~M$_{\odot}$,
which is $\sim$30\% larger than the observed gas mass (Phookun et al. 1993).
To each particle a radius is attributed depending on its mass. 
During the disk evolution the particles can have inelastic collisions, 
the outcome of which (coalescence, mass exchange, or fragmentation) 
is simplified following Wiegel (1994). 
This results in an effective gas viscosity in the disk. 

As the galaxy moves through the intracluster medium (ICM), its clouds are decelerated by
ram pressure. Within the galaxy's inertial system its clouds
are exposed to a wind coming from the opposite direction of the galaxy's 
motion through the ICM. 
The temporal ram pressure profile has the form of a Lorentzian,
which is realistic for galaxies on highly eccentric orbits within the
Virgo cluster (Vollmer et al. 2001).
The effect of ram pressure on the clouds is simulated by an additional
force on the clouds in the wind direction. Only clouds which
are not protected by other clouds against the wind are affected.

To model NGC~4254, the 20\,000 collisional particles with a 1/R column density profile
were added to the 49\,125 non-collisional particles
and the system was evolved during 2~Gyr in order to let it reach 
a relaxed state. At the end of the simulation the difference between the 
total energy and the total angular momentum of the system was smaller 
than 0.5\% of their initial values. During the last Gyr of this simulation
the disk properties (disk scale height, surface density profile, and density profile)
changed less than 5\%. 
The final state of this simulation was used as the initial state for the
simulations presented here.

The companion galaxy with a mass of $10^{11}$~M$_{\odot}$ is simulated by 11\,000 
non-collisional particles forming a Plummer sphere with a core radius of 2~kpc.

\section{The simulations\label{sec:simulation}}

\subsection{The tidal interaction \label{sec:tidal}}

\begin{figure*}
        \resizebox{\hsize}{!}{\includegraphics{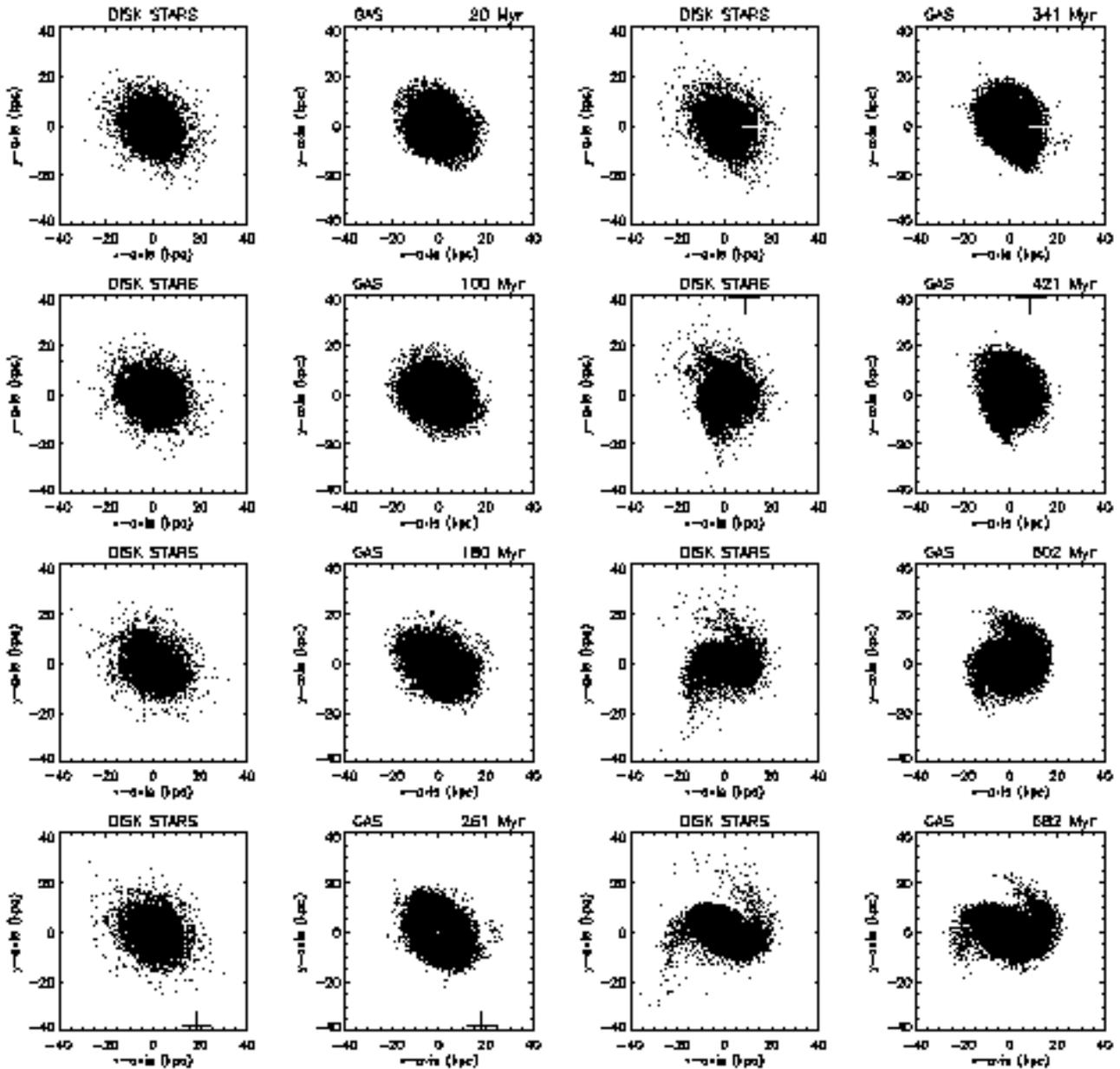}}
        \caption{Snapshots of the evolution of the NGC~4254 system model,
	  undergoing only a tidal interaction with another massive Virgo cluster galaxy.
	  The left panels show the stellar disk, the
          right panels show the gas. The timesteps are marked on the
          upper right of each panel showing the gas.
          The center position of the perturbing galaxy is marked as a cross.
        } \label{fig:evolution165}
\end{figure*}

We have modeled a close and rapid encounter between NGC~4254 and another
massive Virgo cluster galaxy, for which we will try to identify a possible
candidate at the end of this section. The closest distance between the
two galaxies is $D=27$~kpc, which occurs at $t=300$~Myr with a velocity difference
of $710$~km\,s$^{-1}$. Snapshots of this tidal interaction are shown in
Fig.~\ref{fig:evolution165}. The position and inclination angle of NGC~4254
are used. The azimuthal angle is fixed in a way to reproduce the observed
spiral structure. The galaxy rotates clockwise.
NGC~4254 starts building a spiral structure $\sim100$~Myr after the closest passage of
the perturbing galaxy. The spiral arm at the western side, where the interaction
happened, is more prominent than the one at the opposite side. After $t=500$~Myr
an asymmetric spiral pattern is formed with a prominent m=1 mode.

The final location of the perturbing galaxy in our simulation is $9$~kpc east and $120$~kpc north
of the center of NGC~4254, at which point its radial velocity is $-550$~km\,s$^{-1}$ 
with respect to the systemic velocity of NGC~4254. A possible candidate is 
the SB0 galaxy NGC~4262, which is located $51$~kpc east and $137$~kpc north of
NGC~4254. Its radial velocity is $\sim -1050$~km\,s$^{-1}$ with respect to the
systemic velocity of NGC~4254. 
NGC~4262 has a ring of H{\sc i} around it with twice the diameter $D_{25}$ of the optical disk
and an inclination at least 50$^{\circ}$ with respect to the plane of the optical disk
(Krumm et al. 1985).
With an H{\sc i} rotation velocity of $220$~km\,s$^{-1}$ at 2$'$ radius its total 
mass is near the $10^{11}$~M$_{\odot}$ used in our model.
Since NGC~4262 is an S0 galaxy and therefore much more concentrated than a
spiral galaxy, we do not expect to observe an impact of the interaction on
its stellar content (see also NGC~4435 in the NGC~4438/NGC4435 system,
Combes et al. 1988). Concerning the H{\sc i} distribution of NGC~4262,
outer ring structures are frequently observed in gas-rich S0 galaxies, see
e.g. van Driel \& van Woerden (1991), who considered that, in general,
their origins are most likely a galaxy merger or a star formation
burnt-out scenario. Among the S0 galaxies studied by van Driel \& van
Woerden, NGC 4262 has by far the highest inclination difference (60$^{\circ}$)
between the planes of its outer H{\sc i} ring and the stellar disk. For
NGC~4262 we can only speculate that the interaction with NGC~4254 is at
least partly responsible for the peculiar ring structure. Its approximate
symmetry is consistent with the time delay between the gravitational
interaction and today, which is about one rotation period of the ring.

\subsection{Tidal interaction and ram pressure \label{sec:ram}} 

In a second step we introduce a temporal ram pressure profile of the form:
\begin{equation}
p_{\rm ram}=\frac{10\,t_{1}^{2}\,\rho_{0}v_{0}^{2}}{(t-t_{0})^{2}+t_{1}^{2}}\ ,
\end{equation}
where $\rho_{0}=10^{-4}$~cm$^{-3}$, $v_{0}=1000$~km\,s$^{-1}$,
$t_{0}=1$~Gyr, and $t_{1}=115$~My. Since the inclination angle between the disk and the 
galaxy's orbital plane within the cluster
is $70^{\circ}$, the stripping occurs nearly face-on and is very efficient
-- see Vollmer et al. 2001 for a discussion of the stripping efficiency with respect to the
inclination angle. 
While face-on stripping ($i > 60^{\circ}$) and thus a high stripping efficiency is necessary to 
reproduce the observed north-western very low surface brightness plume, 
the inclination angle on the other hand must be smaller than $90^{\circ}$, i.e. face-on, to produce
the north-west asymmetry of the plume. 
For these simulations the shape of the galaxy's orbit within the cluster does not play a dominant role.
The important conditions are (i) that the galaxy is entering the cluster and (ii)
that ram pressure at the time of interest is $p_{\rm ram}(580\ {\rm Myr})=0.7 \rho_{0} v_{0}^{2}$,
consistent with its radial velocity and the ICM density derived from X-rays
(Schindler et al. 1999).

However, when we run our simulations with a constant
cloud column density of $\Sigma_{0}=10^{21}$~cm$^{-2}$, this
ram pressure does not produce a gas tail as observed for NGC~4254,
while increasing the ram pressure to produce such a gas tail has too great an effect
on the gas disk, and pushes a considerable fraction
of the south-eastern part of the galaxy towards the north-west,
which is not observed.

\begin{figure*}
        \resizebox{\hsize}{!}{\includegraphics{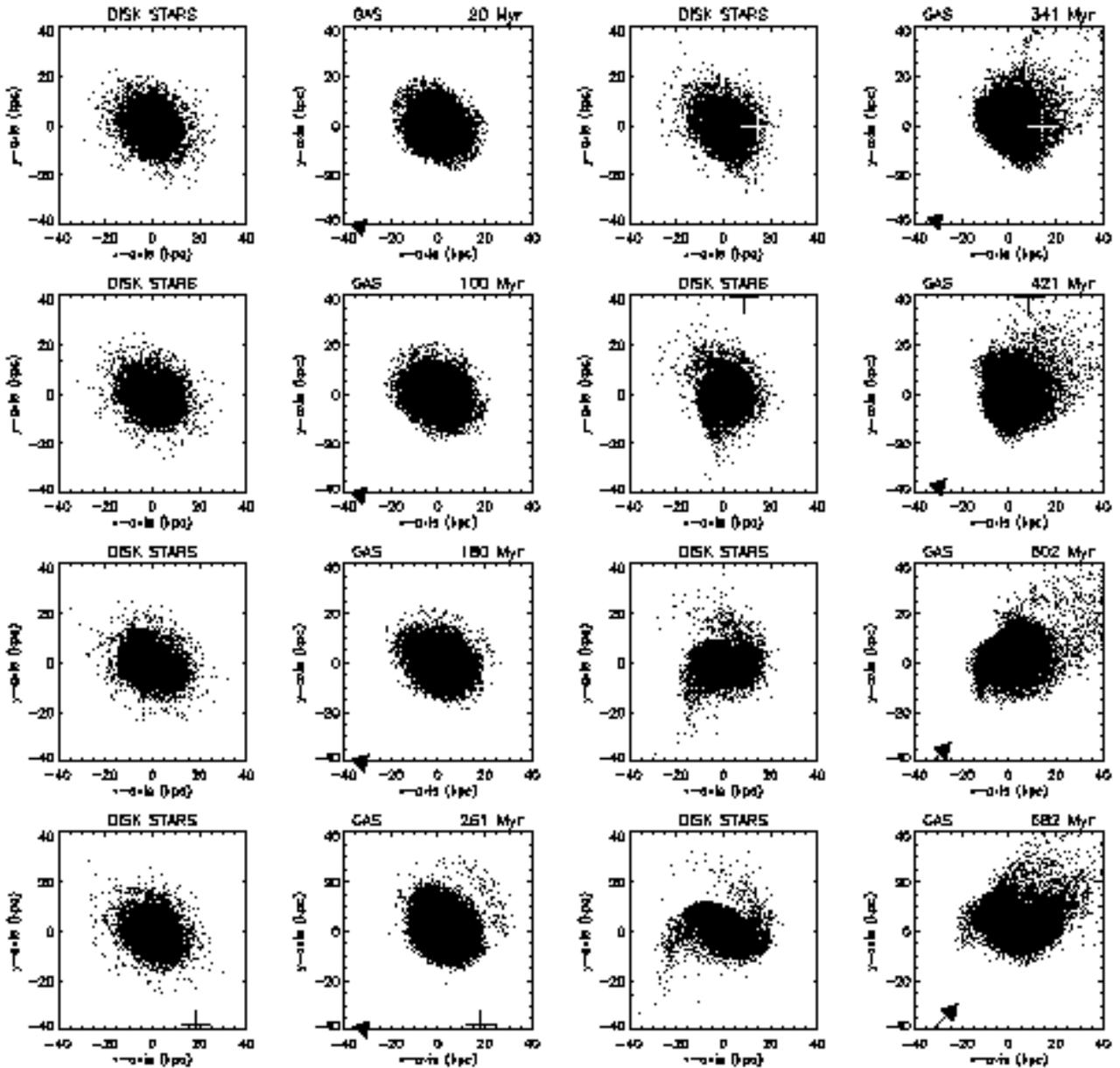}}
        \caption{Snapshots of the evolution of the NGC~4254 system model
	  undergoing as tidal interaction with another massive Virgo cluster galaxy 
	  as well as ram pressure stripping. 
	  The left panels show the stellar disk, the
          right panels show the gas. The timesteps are marked on the
          upper right of each panel showing the gas.
          The center position of the perturbing galaxy is marked as a cross. The arrow indicates
	  the direction of ram pressure, and the length of the arrow is proportional
	  to its strength.
        } \label{fig:evolution204}
\end{figure*}

Therefore, we decided to add the effects of cloud expansion
and evaporation to the numerical code:
once the clouds are pushed out of the galactic disk by ram pressure, they may
expand or evaporate within the hot intra-cluster medium (Cowie \& McKee 1977).
To simulate these effects, the column density of those clouds that are located in regions with 
a local density below a threshold $\rho_{\rm lim}$, is made to decrease exponentially
with time as
$\Sigma=\Sigma_{0}\,\exp(-t/t_{2})$, where $t_{2}=50$~Myr. This corresponds 
approximately to the timescale of saturated evaporation (Vollmer et al. 2001).
The threshold gas density is fixed at $\rho_{\rm lim}=10^{-2}$~cm$^{-3}$, and  
the parameters $t_{2}$ and $\rho_{\rm lim}$ are chosen such that the H{\sc i}
properties of NGC~4254 are reproduced.
Snapshots of the evolution of the NGC~4254 system undergoing a tidal interaction
as well as ram pressure stripping are shown in
Fig.~\ref{fig:evolution204}. The position and inclination angle of NGC~4254
are used.
The evolution of the stellar content is very similar to that in the previous
simulation without ram pressure stripping (Fig.~\ref{fig:evolution204}), because
ram pressure affects selectively only the gas. From the time of closest
passage ($t=300$~Myr) onwards, there is an increase in the difference between
the gas distribution in the tidal interaction simulation and the
mixed interaction simulation.
Since the gas clouds that are expelled from the galaxy by the tidal interaction have
a lower local gas density, their column density decreases, and they are stripped
easily by a moderate amount of ram pressure. This can be noted by the amount of
low volume density gas that has been pushed towards the north-west at $t=400$~Myr.
In the forming western spiral arm the high local gas density 
prevents the gas from being stripped, and only the gas located outside the arm is stripped.

\subsection{Comparison between observations and simulations \label{sec:compare}}

\begin{figure}
        \resizebox{10cm}{!}{\includegraphics{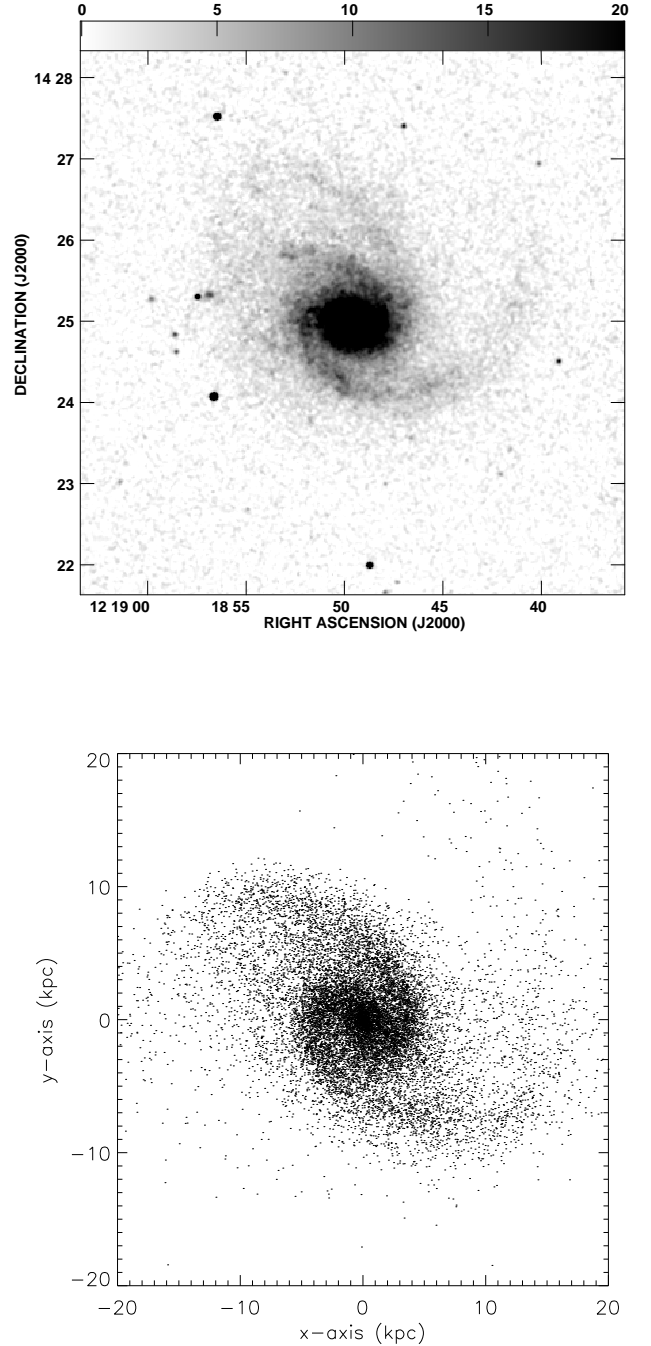}}
        \caption{The stellar component of NGC~4254. Upper panel: 2MASS $K$ band image (from NED).
	  Lower panel: model distribution, for the model with both tidal interaction
	  and ram pressure stripping; snapshot at $t=580$~Myr.
        } \label{fig:stars}
\end{figure}

We chose the snapshot
at $t=580$~Myr for the comparison with the observations, because (i) it reproduces
the spiral pattern of NGC~4254 and (ii) it shows a very low averaged surface density
tail to the north-west as it is observed. At that moment the ram pressure is only 
$p_{\rm ram}=0.7 \rho_{0} v_{0}^{2}=70$~cm$^{-3}$(km\,s$^{-1}$)$^{2}$.
Using the Gunn \& Gott (1972) approximation we obtain:
\begin{equation}
\Sigma v_{\rm rot}^{2} R^{-1} = p_{\rm ram}\ ,
\end{equation}
where $\Sigma$ is the gas surface density, $v_{\rm rot}$ is the rotation velocity
of the galaxy and $R$ is the galactic radius.
With $v_{\rm rot}=170$~km\,s$^{-1}$ and $\Sigma=10^{20}$~cm$^{-2}$ we obtain
$R=14$~kpc, or about the optical size $D_{25}$ of the galaxy.
Thus ram pressure affects only gas with $\Sigma < 10^{20}$~cm$^{-2}$.
Since the radius where the radially averaged $\Sigma=10^{20}$~cm$^{-2}$ of the H{\sc i}
gas in NGC~4254 is 18~kpc (Warmels et al. 1988, Cayatte et al. 1994), i.e.
1.2 times the optical radius, we do not expect that the galaxy would be affected by
ram pressure. We suggest that the peculiar low surface density gas distribution
and kinematics are due to the interplay between the tidal interaction and
ram pressure stripping. The observed and the simulated distribution of the stellar content
are shown in Fig.~\ref{fig:stars}.
Like the observed main spiral pattern, the simulated spiral structure is asymmetric with a wider, 
western spiral arm
and a tighter, eastern spiral arm. In addition, the observed spiral structure shows an m=3 and an 
m=5 mode, which are not reproduced by the model. The excitation of these modes depends crucially
on the mass distribution within the galaxy. Since the aim of this study is not to reproduce
exactly the complicated spiral pattern of NGC~4254 but its gas distribution and kinematics,
we have not attempted to investigate the influence of different mass distributions on
the spiral pattern, which also would require a larger number of particles.

\subsection{The moment maps \label{sec:moments}}

\begin{figure}
        \resizebox{8cm}{!}{\includegraphics{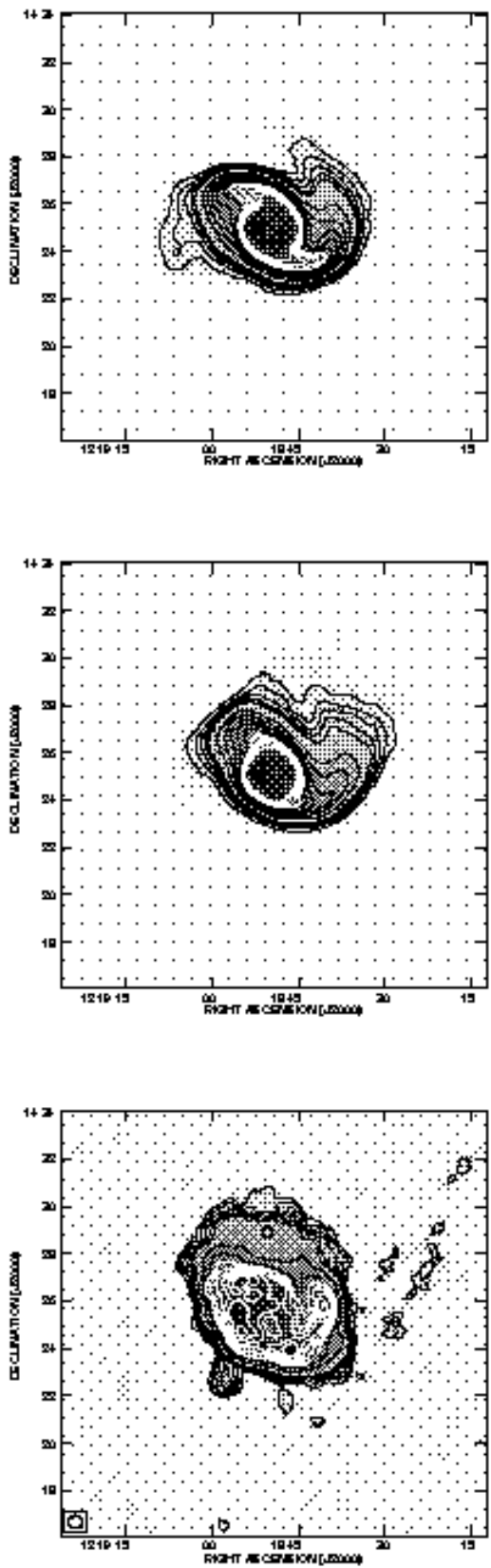}}
        \caption{The gaseous component of NGC~4254.
	Upper panel: model distribution of the H{\sc i} gas
        for the model with only a gravitational interaction.
        Middle panel: model distribution of the H{\sc i} gas
        for the model with gravitational interaction plus 
        ram pressure stripping.
        Lower panel: H{\sc i} distribution of NGC~4254 observed by Phookun et al. (1993).
	The contour levels in all images are (0.02, 0.04, 0.06, 0.08, 0.1, 0.2, 0.3, 0.4, 0.5,
	0.6, 0.7, 0.8, 0.9, 1.0) times (i) half of the maximum of the image for the models
	and (ii) the maximum of the image for the H{\sc i} data.
        } \label{fig:mom0}
\end{figure}
For the moment maps of the H{\sc i} data we smoothed the cube to a resolution of 
$35'' \times 35''$ and clipped the data at a 0.8~mJy/beam/channel level.
A comparison between simulated and observed gas distributions at $t=580$~Myr
is shown in Fig.~\ref{fig:mom0}. Since we do not differentiate between atomic and
molecular hydrogen, the simulated gas distribution shows the total gas content.
Although the model gas distributions are more concentrated in the central part of the
galaxy than is observed in H{\sc i}, it should be noted that in the central few kpc
of NGC~4254 most of the gas is molecular (Kenney \& Young 1988), which makes the
total gas distribution resemble the simulations.
In the simulations without ram pressure stripping, the gas follows closely the stellar
spiral structure, as expected. This is also the case in the inner parts of the galaxy when
ram pressure stripping is included, whereas the outer parts of the gas distribution show
important differences, since
ram pressure has pushed the gas from the south-eastern quadrant towards
the center. We suggest that the observed local H{\sc i} maximum in the south-east
is due to this effect. 
The low surface density northern extension is fairly well reproduced 
by the simulations including ram pressure stripping. However, this feature
extends more to the west in the observed data. The observed north-western, very low
surface density plume is present in the simulations including ram pressure stripping,
but their surface density structures differ. Whereas in the simulations there is a gradient to
lower surface densities to the north, the observations show a sudden drop 
between the galactic main disk and the plume (see also Fig.~\ref{fig:deep}). 
Within the plume the surface density seems to be rather constant. 
The observed southern low surface density blob is not present in our simulations.

\begin{figure}
        \resizebox{8cm}{!}{\includegraphics{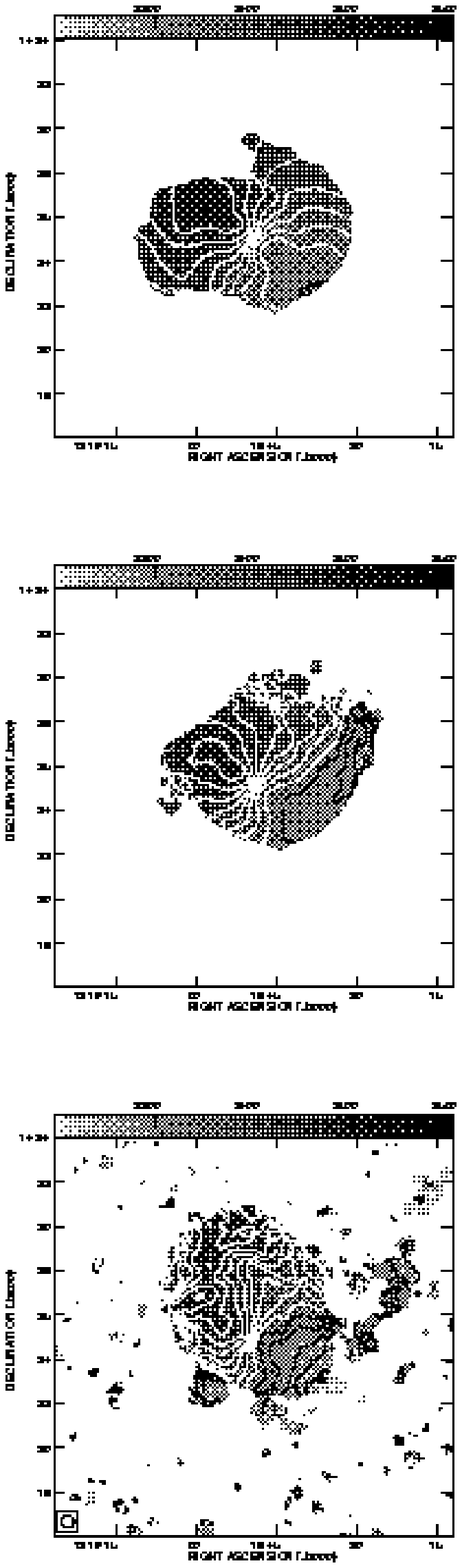}}
        \caption{Top panel: model distribution of the H{\sc i} velocity field
        for a model of a gravitational interaction alone.
        Middle panel: distribution of the H{\sc i} velocity field
        of the model of a gravitational interaction together with 
        ram pressure.
        Bottom panel: H{\sc i} velocity field of NGC~4254 (Phookun et al. 1993).
	The contour levels of all images are at 2250-2510~km\,s$^{-1}$, at
	intervals of $20$~km\,s$^{-1}$.
        } \label{fig:mom1}
\end{figure}
A comparison between the simulated and observed velocity fields at $t=580$~Myr is shown
in Fig.~\ref{fig:mom1}.
The effects of ram pressure stripping on the velocity field are the following:
(i) on the eastern side of the galaxy the gas with low radial velocities is
 pushed towards the north-west, and some of the gas is decelerated towards
the cluster mean velocity (1100~km\,s$^{-1}$). The net effect is that the eastern
maximum is displaced to the south and has a more narrow shape;
(ii) the western plateau of the rotation curve is less pronounced when ram pressure
stripping is added to the simulations; (iii) the velocity contours are straightened
to the north, but the imprint of the spiral arm's velocity structure is still visible.
This velocity structure is more pronounced in the
observed velocity field. The straightening of the model velocity contours to the north
can be observed in Fig.~\ref{fig:deep}.
The observed negative velocities with respect to the systemic velocity at
RA=12$^{\rm h}$18$^{\rm m}$50$^{\rm s}$ and DEC=14$^{\circ}$28$'$00$''$ 
do not have counterparts in the simulations.

\begin{figure}
        \resizebox{8cm}{!}{\includegraphics{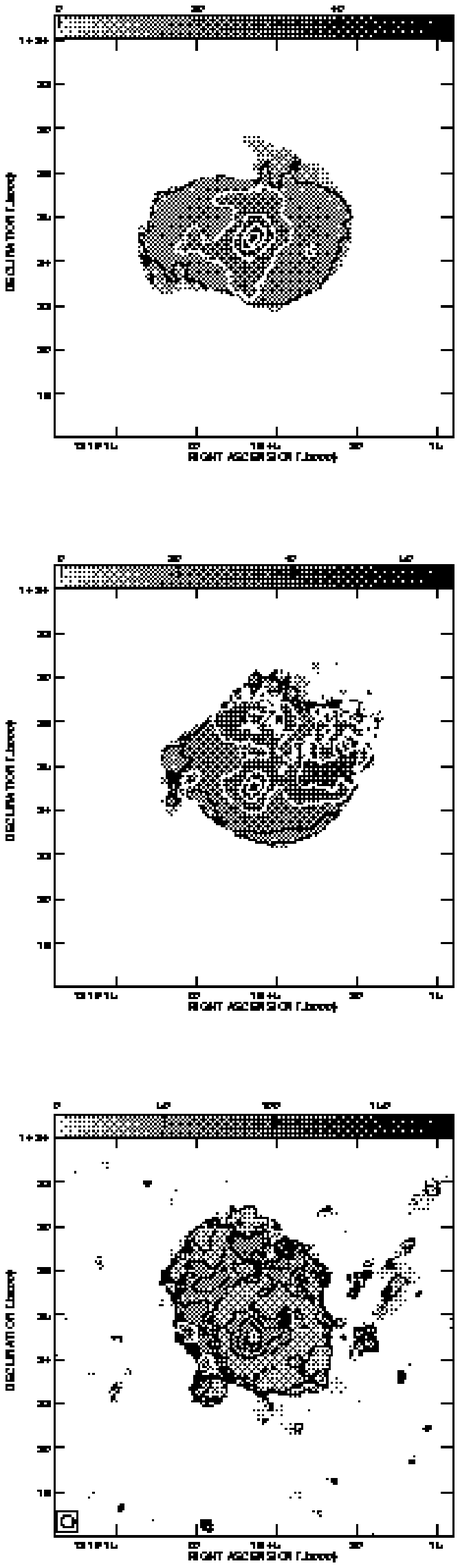}}
        \caption{The gaseous velocity dispersion in NGC4254.
	Upper panel: model gas velocity dispersion map
        for the model with only a gravitational interaction.
        Middle panel: model gas velocity dispersion map
        for the model with both gravitational interaction and 
        ram pressure stripping.
        Lower panel: H{\sc i} velocity dispersion map (Phookun et al. 1993).
	The contour levels in all images are at 10-60~km\,s$^{-1}$, in steps
	of $10$~km\,s$^{-1}$.
        } \label{fig:mom2}
\end{figure}

The comparison between simulated and observed velocity dispersions at $t=580$~Myr are shown in
Fig.~\ref{fig:mom2}.
The velocity dispersion of the simulations without ram pressure stripping show only a
significant enhancement in the galaxy center, as expected.
When ram pressure stripping is included, three additional local maxima are found:
(i) to the east, (ii) to the north, and (iii) to the west of the center.
The observed velocity dispersions also show extended maxima to the east, north andwest
of the galaxy center (Fig.~\ref{fig:deep}).

\subsection{The three dimensional data \label{sec:3D}}

Since the three dimensional H{\sc i} distribution of NGC~4254 has a complex structure, 
moment maps do not show all the information contained in the data cubes.
To complete the information contained in the moment maps, 
we show some selected position-velocity diagrams as well as
a three dimensional visualisation of the model and observed H{\sc i} cubes.

\begin{figure*}
        \resizebox{\hsize}{!}{\includegraphics{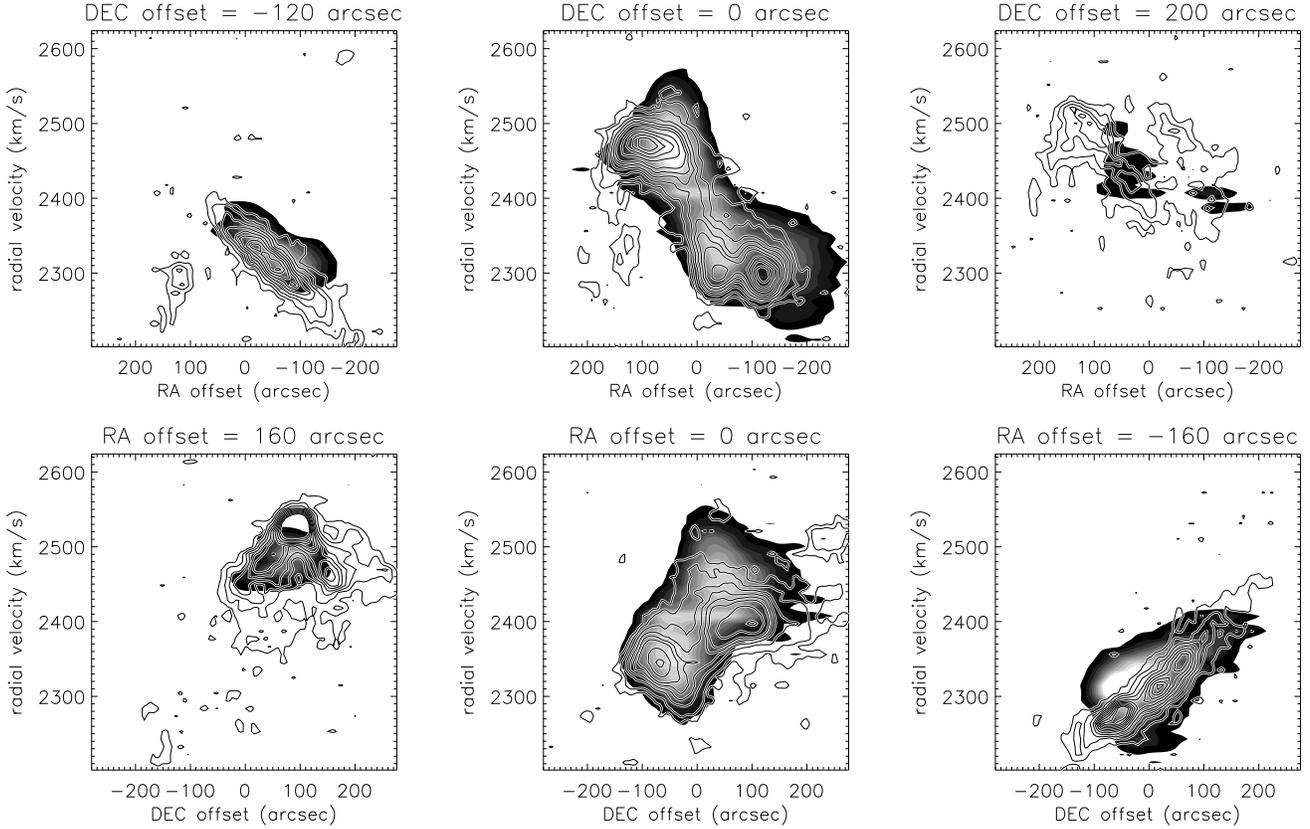}}
        \caption{Position-velocity diagrams of 6 selected planes.
	  Contours: H{\sc i} observations (Phookun et al. 1993).
	  Greyscale: model including ram pressure. Dark regions correspond
	  to regions of low intensity.
        } \label{fig:pv}
\end{figure*}

Fig.~\ref{fig:pv} shows the H{\sc i} data and the model including a gravitational
interaction and ram pressure in 3 position-velocity diagrams along the declination
axis and 3 others along the right ascension axis. The H{\sc i} data
is represented by contours and the model by the greyscale. The observed main features are all
qualitatively reproduced. 
However, the low velocities of the low intensity, south-eastern 
edge ($\Delta \delta = -120'',\ 0''$; $\Delta \alpha = 160''$) are not present in the model. 
We speculate that this represents gas initially located at the tip of the spiral arm
that is decelerated by ram pressure to smaller velocities. 
However, a captured dwarf scenario might also be possible, which affects the gas
kinematics locally. No optical counterpart is detected at this location.
In addition, the model gas distribution at the northern edge ($\Delta \delta = 200''$) is 
offset from the observed gas distribution as it can be seen in Fig.~\ref{fig:mom0}.

\begin{figure}
        \resizebox{8cm}{!}{\includegraphics{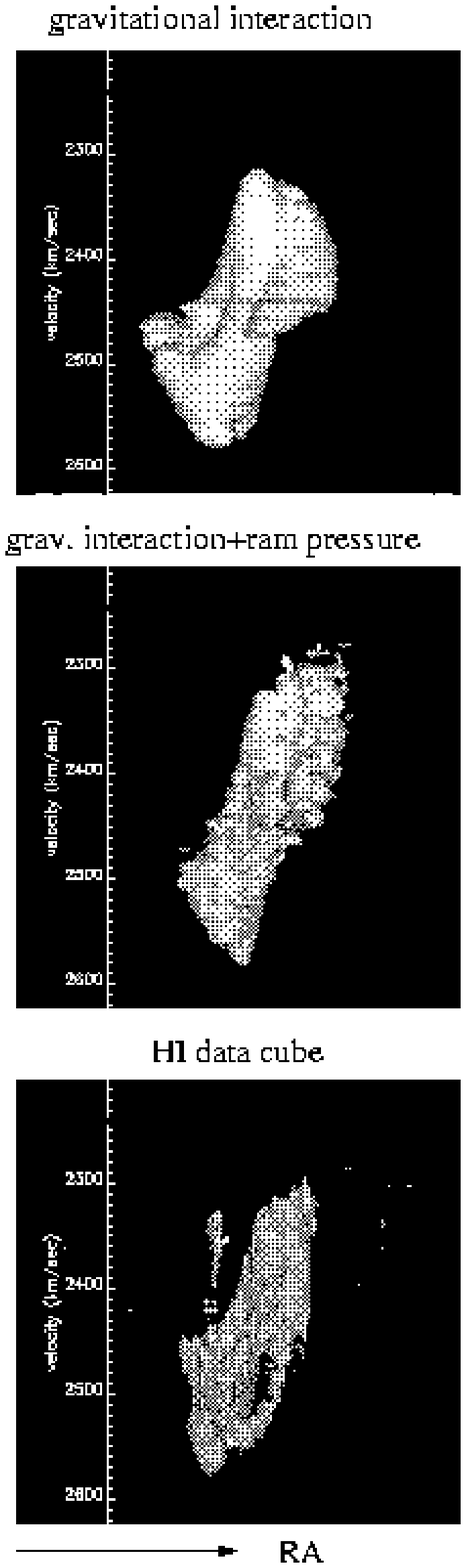}}
        \caption{3D representation of data cubes. 
	  The axis of right ascension is indicated. The axis of declination
	  is perpendicular to the image plane. Upper panel: model cube
	  of the simulation including a gravitational interaction alone.
	  Middle panel: model cube of the simulations including a
	  gravitational interaction and ram pressure. Lower panel: observed
	  H{\sc i} data cube (Phookun et al. 1993). The intensity isosurface is at
	  2~mJy/beam/channel.
        } \label{fig:3D}
\end{figure}

A three dimensional visualisation of the full data set
is shown in Fig.~\ref{fig:3D}, where the upper and middle panels represent,
respectively, the final snapshots
of the simulations (i) with a gravitational interaction alone (cf. Fig.~\ref{fig:evolution165})
and (ii) with a gravitational interaction and ram pressure (cf .Fig.~\ref{fig:evolution204}). 
The lower panel shows the H{\sc i} data cube of Phookun et al. (1993)
with an isosurface at 2~mJy/beam/channel.
The two tidally induced spiral arms are clearly visible to the left and right of
the upper panel. By comparing the upper and lower panel the effects of ram pressure
on the model gas distribution can be seen directly: the extent of the gas distribution
in right ascension is decreased (cf. Fig.~\ref{fig:mom0}), the velocities in the
high velocity arm are increased (the gas is accelerated; the wind and galactic rotation
have the same direction) and those in the low velocity arm are decreased
(the gas is decelerated; the wind and galactic rotation have opposite directions).
These features can be qualitatively recognised in the H{\sc i} data cube
(Fig.~\ref{fig:3D} lower panel). In addition, the quasi horizontal edge at
$v \sim 2460$~km\,s$^{-1}$ to the right of Fig.~\ref{fig:3D} (upper panel) is
stretched to higher velocities in  Fig.~\ref{fig:3D} (middle panel) due to the
gas acceleration by ram pressure. However, we note that the observed gas velocities in
Fig.~\ref{fig:3D} (lower panel) at that location in the data cube are shifted to
much higher velocities ($\Delta v \sim 100$~km\,s$^{-1}$). 
Moreover, the low velocities in the H{\sc i} blob located in the upper left
quadrant of Fig.~\ref{fig:3D} (lower panel), which corresponds to the southern blob 
of Fig.~\ref{fig:mom0}, are not reproduced. If we lower the intensity level of
Fig.~\ref{fig:3D} (middle panel), single gas clouds become visible in this region
of the cube, but they do not form an elongated structure as it is observed.

Thus we conclude that although the observed main features are reproduced qualitatively,
there are quantitative differences between observations and simulations.
However, we note that effects of ram pressure definitely improve the fit to
the H{\sc i} observations (gas displacement, acceleration and deceleration).
Only the amplitude of the mechanism is not strong enough to reproduce the
H{\sc i} in the lower right and upper left quadrant of Fig.~\ref{fig:3D} (lower panel).
We speculate that this is mainly due to the missing physics of H{\sc i} cloud expansion and 
evaporation in the model and, in a less important way, to the amount of ram
pressure and the inclination angle between the disk and the galaxy's orbital plane
within the cluster.

\subsection{The spectra \label{sec:compspec}}

\begin{figure}
        \resizebox{8cm}{!}{\includegraphics{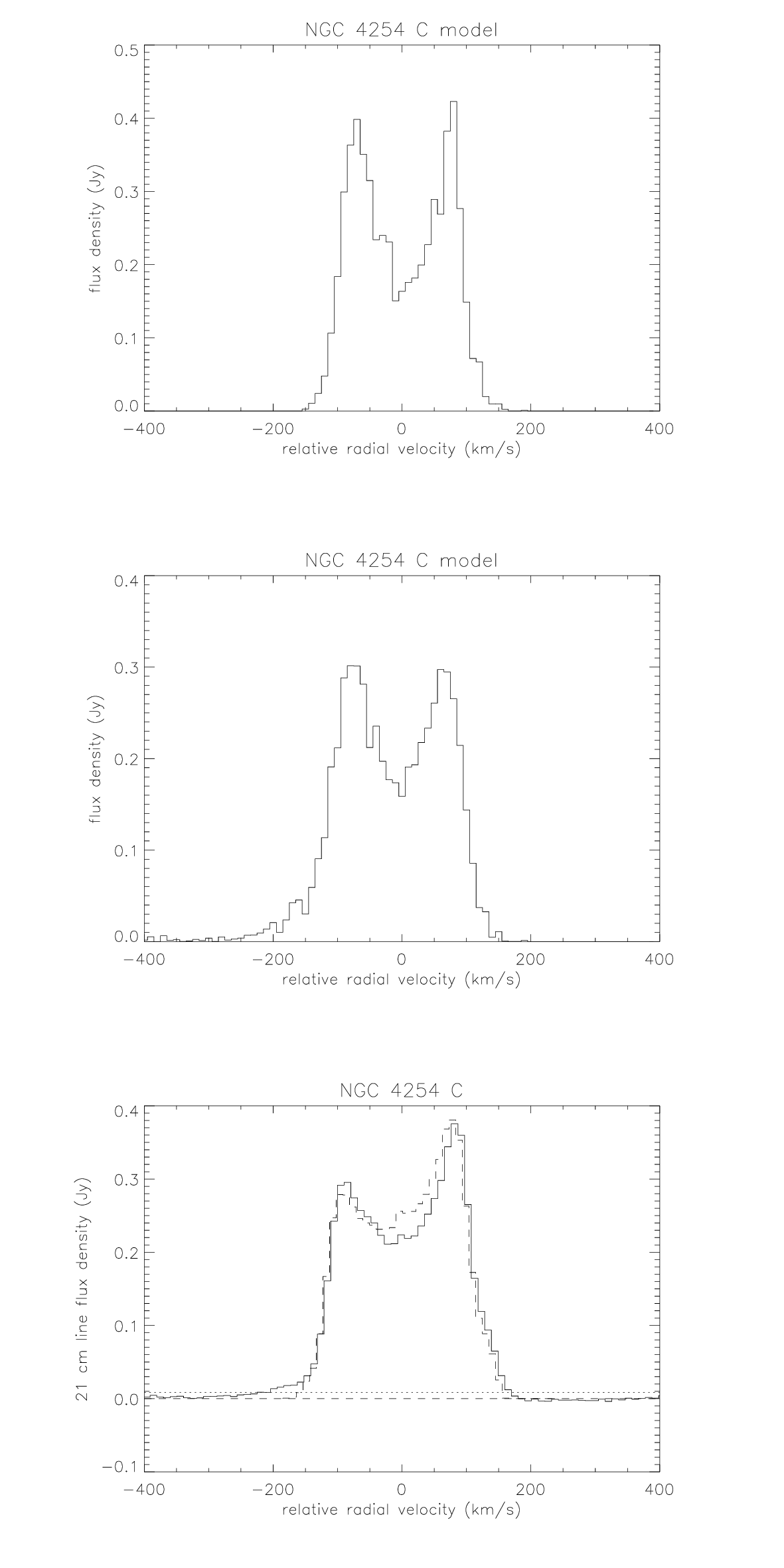}}
        \caption{The central H{\sc i} spectrum of NGC~4254.
	Upper panel: Simulated central H{\sc i} spectrum at $t=580$~Myr for the model with only a 
	gravitational interaction.
	Middle panel: Simulated central H{\sc i} spectrum at $t=580$~Myr of the model with both
	gravitational interaction and ram pressure stripping.
	Lower panel: Observations. Solid lines: Effelsberg 100-m single dish spectra of the 
	four off-center positions.
	Their locations with respect to the galaxy center are marked on top of each panel.
	Dashed line: synthesized VLA spectrum. Dotted line: 3$\sigma$ noise level of the
	100-m spectrum. Radial velocities are given relative to the systemic velocity of NGC~4254.
        } \label{fig:cspectra}
\end{figure}
The observed and simulated H{\sc i} spectra of the main body of the H{\sc i} distribution
(see Sect.~\ref{sec:results}) are shown in Fig.~\ref{fig:cspectra}.
While all spectra show the characteristic double-horned structure, only the simulation
that includes ram pressure stripping reproduces the observed tail at low velocities.

\begin{figure}
        \resizebox{\hsize}{!}{\includegraphics{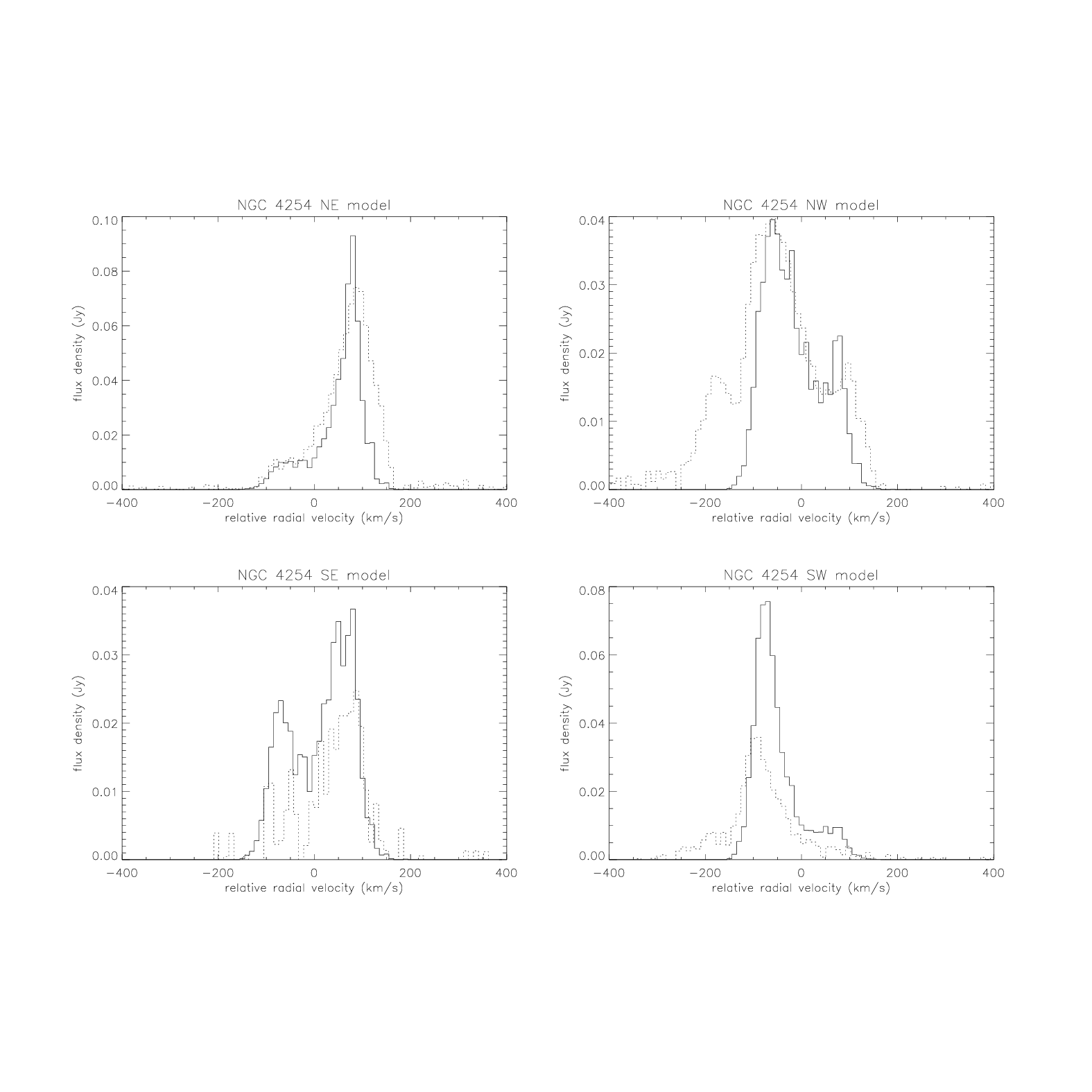}}
        \caption{Off-center positions H{\sc i} spectra of NGC~4254.
	  Solid line: simulated off-center spectra at $t=580$~Myr for the model with only a gravitational 
	  interaction. Dotted line: observed Effelsberg off-center spectra.
	  The locations of the spectra with respect to the galaxy center (see Sect.~\ref{sec:results})
	  are marked on top of each panel. Radial velocities are relative to the systemic
	  velocity of NGC~4254.
        } \label{fig:spectra165}
\end{figure}
\begin{figure}
        \resizebox{\hsize}{!}{\includegraphics{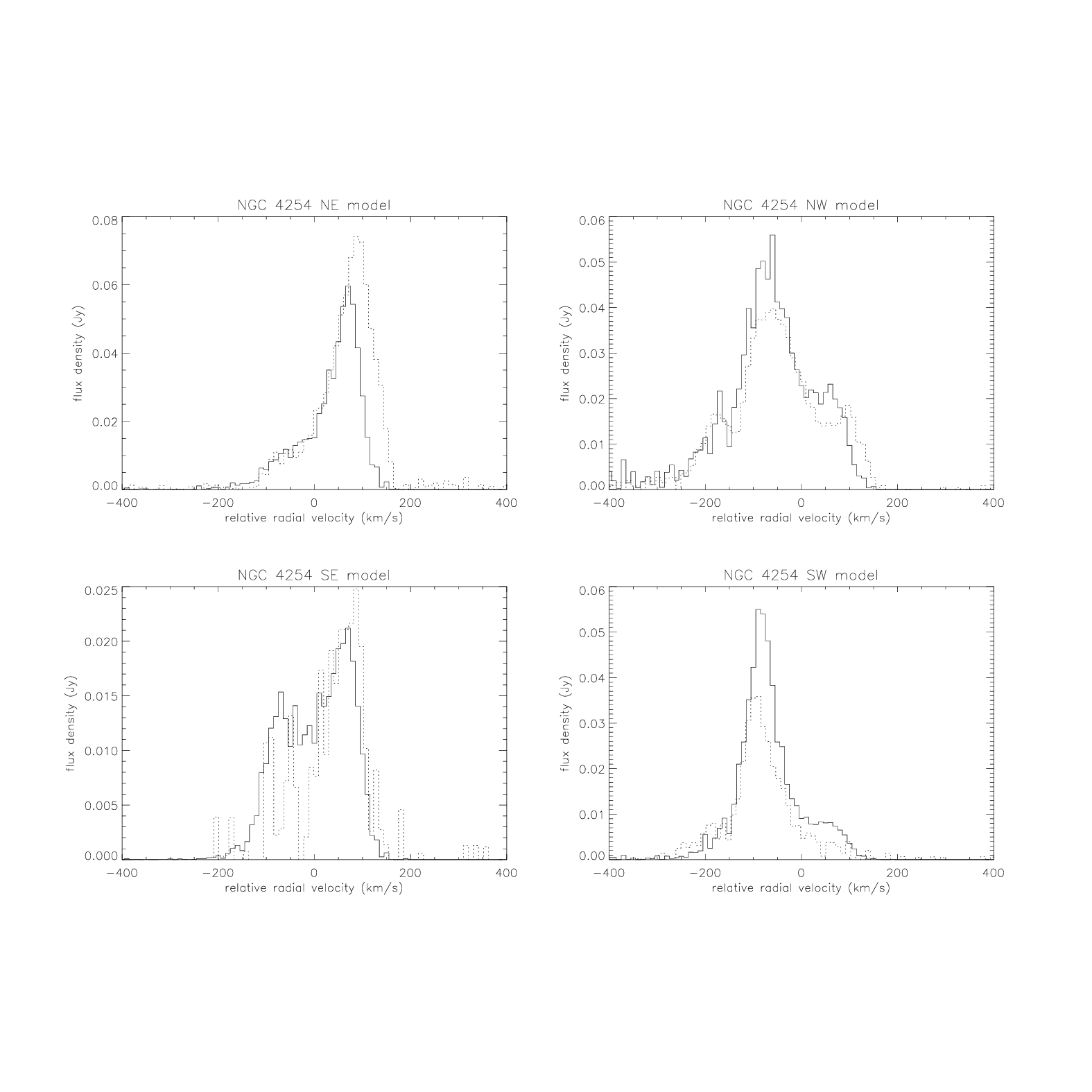}}
        \caption{Off-center positions H{\sc i} spectra of NGC~4254.
	  Solid line: simulated off-center spectra at $t=580$~Myr for the model with both gravitational 
	  interaction and ram pressure stripping. Dotted line: observed Effelsberg off-center spectra.
	  The locations of the spectra with respect to the galaxy center (see Sect.~\ref{sec:results})
	  are marked on top of each panel. Radial velocities are relative to the systemic
	  velocity of NGC~4254.
        } \label{fig:spectra204}
\end{figure}
The observed and simulated H{\sc i} spectra of the 4 off-center positions (see Sect.~\ref{sec:results})
are shown in Fig.~\ref{fig:spectra165} for the model with only gravitational interaction and
in Fig.~\ref{fig:spectra204} for the model with both gravitational interaction and
ram pressure stripping.

In the model without ram pressure stripping the spectrum at the NE position can already be
reproduced well. The shift of the observed H{\sc i} spectrum to higher velocities
at $v = 100 - 150$~km\,s$^{-1}$ is due to the fact that the model does not reproduce the
high velocities of the northern H{\sc i} arm (see Sect.~\ref{sec:3D}).
Although the shape of the SE spectrum is reproduced, the observed flux densities
are 1/3 higher than the simulated ones. The simulated SW spectrum lacks the observed
emission at low velocities, and the observed and simulated flux densities differ by a factor of two.
The simulated NW spectrum completely lacks the observed peak at low velocities.

In the model including ram pressure stripping the simulated NE spectrum is almost unchanged and 
reproduces well the observations.
The simulated SE and SW spectra have lower flux densities than those for the simulations without ram
pressure (Fig.~\ref{fig:spectra165}) and give better fits to the observed spectra.
The effects of ram pressure stripping on the NW spectrum are twofold:
(i) the high-velocity maximum is decreased and (ii) a low velocity tail appears -- both effects
are observed. Thus all 5 observed spectra are well, or, in some cases better reproduced 
by the simulations including ram pressure stripping.

\section{Discussion \label{sec:discussion}}

Previously, two explanations were proposed for the m=1 mode of NGC~4254: (i) its intrinsic mass distribution
that favors the formation of this spiral pattern (Iye et al. 1982), and (ii) the excitation 
of the m=1 mode by infalling H{\sc i} gas together with a swing amplifier (Phookun et al. 1993). 
Based on our simulations we suggest another mechanism: a close and rapid tidal interaction 
that triggers an m=1 mode. Our models indicate that NGC~4254,
when entering the Virgo cluster, had a close and rapid encounter 
with a $10^{11}$~M$_{\odot}$ galaxy $\sim 280$~Myr ago, and has since then been undergoing increasing
ram pressure stripping. The tidal interaction caused the m=1 spiral structure, and, together,
the tidal interaction and ram pressure caused the peculiar H{\sc i} distribution and kinematics.
The intrinsic mass distribution of NGC~4254 then favors the formation of additional m=3 and m=5 spiral structures.

Phookun et al. (1993) discarded the ram pressure scenario on the following grounds:
since NGC~4254 is moving towards the south-east,
which makes the ram pressure push the gas located in the south-eastern part of the galaxy towards the north-west,
there can be no gas ahead of the galaxy as it moves into the intracluster medium.
However, the observed low surface density gas blobs to the south of the main H{\sc i} body represent such gas.

Although it is true that in an edge-on stripping scenario the existence of the southern,
low surface density gas blobs would rule out the ram pressure scenario,
in our model stripping occurs almost face-on. In the case of a face-on stripping the
gas is removed in a cone-like structure and
the projected radial distance of the stripped gas increases with its height from the disk
(see Vollmer \& Huchtmeier 2003).
This makes it in principle possible to observe low surface density gas
seen in projection to the south of the main gas disk, which represents
stripped gas that is actually located in front of NGC~4254 and which should show negative velocities
with respect to the systemic velocity of NGC~4254 as is observed.
The southern, low surface density gas blobs are thus not contradictory to a
face-on ram pressure stripping scenario, and we in fact find gas clouds in the south of the gas disk in the
model snapshot that we compare to the observations in Fig.~\ref{fig:evolution204} -- these
are not visible in the gas distribution (Fig.~\ref{fig:mom0}) and 3D visualisation
(Fig.~\ref{fig:3D}), because their column density is very low. 

Although our model including ram pressure stripping succeeds in reproducing the spiral structure 
and the overall gas distribution and kinematics, 
the main difference between our model and the observations is the shape of the extended,
low surface density gas tail observed to the north-west. Whereas the simulated gas tail 
is a smooth continuation of the m=1 spiral arm, the observed tail consists of single blobs, at the 
level of sensitivity of the VLA data, and there is a gap between the gas disk and the tail,
which might have two reasons:
(i) Once the stripped gas leaves the disk it is ionized and expanding through ICM-ISM mixing, 
and is thus no longer visible in the H{\sc i} line. At a later stage it might cool and recombine again.
(ii) The stripping is not continuous, but occurs in discrete events, because the ICM
is not homogeneous. In this case the observed tail might represent gas falling back.
We did not attempt to simulate this effect, because we do not have enough information
on the temporal profile of such discrete events.

The evolutionary history of NGC~4254 thus closely resembles that of NGC~4654 (Vollmer 2003).
Both galaxies had a close and rapid encounter with another massive Virgo cluster galaxy and
both galaxies are presently still undergoing ram pressure stripping. In both cases, the tidal interactions
make the stripping more efficient. For NGC~4254 we had to include the effects of
gas expansion and/or evaporation into the simulations, which was not necessary for NGC~4654 --
if we had included these effects in the NGC~4654 simulations, we could have used a
ram pressure that increases with time and whose maximum value would have been a few times smaller.

Recently Minchin et al. (2005) detected $1.6\times 10^{8}$~M$_{\odot}$ of
neutral hydrogen in the object VIRGOHI21 which has a radial velocity of
$\sim 2000$~km\,s$^{-1}$ and lies $\sim 120$~kpc north-west from NGC~4254.
The total H{\sc i} flux of VIRGOHI21 is too low and its position too far away
from the beam center of our NW position to be detectable in the Effelberg
spectrum. We exclude that VIRGOHI21 represents tidal debris from NGC~4254
for two reasons: (i) the velocity gradient within VIRGOHI21 is opposite to
the gradient between NGC~4254 and VIRGOHI21 and (ii) if VIRGOHI21 were
tidal debris, the second interacting galaxy should be located roughly on
an extrapolated line between NGC~4254 and VIRGOHI21, but there is no
galaxy close enough in velocity in this region of the sky.

\section{Conclusions \label{sec:conclusions}}

NGC~4254 is a peculiar spiral galaxy located at the periphery of the Virgo cluster.
It has a prominent m=1 spiral pattern, making it look basically one-armed, 
and its H{\sc i} distribution and kinematics
are quite distorted. Deep VLA H{\sc i} imaging (Phookun et al. 1993) showed kinematically perturbed, 
low surface density gas around the galaxy, in particular, 
an extended, very low surface density tail to the north-west of the galaxy.
We observed NGC~4254 in H{\sc i}
with the Effelsberg 100-m telescope at 5 different positions: 
towards the center and at 4 positions located one 9.3$'$ HPBW to the NE, NW, SW, and SE from the center.
Our NW spectrum shows $\sim30$\% more low surface density gas than the VLA data, and
a bimodal velocity distribution in the part that corresponds to the north-western gas tail.

We investigated if a scenario involving a tidal interaction and subsequent ram pressure stripping
could explain these properties.

In a first step we modeled a close and rapid encounter between NGC~4254 and another
massive galaxy. In this way we were able to reproduce the observed asymmetric spiral structure
$\sim 280$~Myr after the passage of the perturbing galaxy, for which 
we identified NGC~4262 as a potential candidate.

In a second step we included the effects of ram pressure stripping that slowly increases with time
since we assume that NGC~4254 is entering the Virgo cluster, in an almost face-on stripping
geometry, with an angle of $70^{\circ}$ between the galaxy disk and the orbital plane.
At its location at the periphery of the Virgo cluster, the small ram pressure 
($p_{\rm ram} \sim 100$~cm$^{-3}$(km\,s$^{-1}$)$^{2}$) would not cause significant H{\sc i} stripping
of NGC~4254 without the tidal interaction event that has pulled out gas from the disk.
The observed H{\sc i} properties
can only be reproduced if we include the effects of gas expansion and/or evaporation into 
our simulations. Combining all effects we succeeded in reproducing the main characteristics of the
H{\sc i} spectra, gas distribution, and velocity field, as well as the m=1 spiral pattern. 

Thus, NGC~4254 had an evolutionary history in the Virgo cluster similar to that of NGC~4654 
(Vollmer 2003);
both are entering the Virgo cluster for the first time and experienced a tidal
interaction and still undergo ram pressure stripping.

\begin{acknowledgements}
Based on observations with the 100-m telescope of the MPIfR (Max-Planck-Institut f\"{u}r 
Radioastronomie) at Effelsberg.
BV would like to thank L. Hernquist for making his program that
generates initial conditions available and
the NCSA Astronomy Digital Image Library (ADIL) for providing the VLA data cube 
for this article. This research has made use of the NASA/IPAC Extragalactic Database 
(NED) which is operated by the Jet Propulsion Laboratory, California Institute
of Technology, under contract with the National Aeronautics and Space Administration.

We would like to thank the anonymous referee his useful comments which helped to
improve this article.
\end{acknowledgements}

\end{document}